\documentclass[aps,prl,twocolumn]{revtex4}

\usepackage{epsfig}
\usepackage{graphicx}  
\usepackage{bm}        
\usepackage{amssymb}   

\begin{document}

\title{ Fabrication and characterization of large arrays of mesoscopic gold rings on  large-aspect-ratio cantilevers
}

\author{ D.Q. Ngo$^{1}$, I. Petkovi\'{c}$^{1}$, A. Lollo$^{1}$, M.A. Castellanos-Beltran$^{2}$, and
J.G.E. Harris$^{1,3}$
}

\affiliation{
 $^{1}$  Department of Physics, Yale University, New Haven, Connecticut 06520, USA\\
 $^{2}$ National Institute for Standards and Technology, Boulder, CO, 80305, USA\\
 $^{3}$ Department of Applied Physics, Yale University, New Haven, Connecticut 06520, USA \\
    }

\email{ivana.petkovic@yale.edu}

\begin{abstract}
We have fabricated large arrays of mesoscopic metal rings on ultrasensitive cantilevers. The arrays are defined by electron beam lithography and contain up to $10^5$ rings. The rings have a circumference of 1 $\mu$m, and are made of ultrapure (6N) Au that is deposited onto a silicon-on-insulator wafer without an adhesion layer. Subsequent processing of the SOI wafer results in each array being supported at the end of a free-standing cantilever. To accommodate the large arrays while maintaining a low spring constant, the cantilevers are nearly 1 mm in both lateral dimensions and 100 nm thick. The extreme aspect ratio of the cantilevers, the large array size, and the absence of a sticking layer are intended to enable measurements of the rings' average persistent current $\langle I \rangle$ in the presence of relatively small magnetic fields. We describe the motivation for these measurements, the fabrication of the devices, and the characterization of the cantilevers' mechanical properties. We also discuss the devices' expected performance in measurements of $\langle I \rangle$.
\end{abstract}

\maketitle

\email{ ivana.petkovic@yale.edu}

\section{Introduction}

An isolated conducting ring in its quantum ground state (or, at finite temperature, in thermal equilibrium) supports a non-zero current whenever the magnetic flux $\phi$ that it encloses is a non-integer multiple of the flux quantum $\phi_0 = h/e$ \cite{bloch65,bloch68,schick68,gunther69, buttiker83,landauer_buttiker85,cheung_shih}. This equilibrium current is known as persistent current, and reflects the sensitivity of the electron wavefunction phase to the ring's topology and the flux $\phi$. As a result, persistent current can only be observed if the electron phase coherence length is greater than the ring's circumference.

Persistent current has been a topic of considerable interest since it was first predicted to be observable in microfabricated rings \cite{cheung_imb89,cheung_riedel_gefen_prl62_89,montambaux90,riedel_vonoppen93}. In disordered metal rings with circumference $\sim$ 1 $\mu$m its magnitude  is predicted to be roughly $\sim$ 1 nA, and to be suppressed on a temperature scale $\sim$ 1 K \cite{riedel_vonoppen93}. The current exists only within the closed ring, and so cannot be measured using conventional ammeters. Instead, experiments have mostly inferred it by measuring its magnetic moment (or related quantities). This presents a substantial technical challenge, as the magnetic moment produced by the persistent current in such a ring is typically $\sim 10 \, \mu_B$.

Persistent current can be distinguished from many background signals because it oscillates as a function of $\phi$ with period $h/e$. More specifically, in an isolated disordered metal ring it is given by

\vspace{-1mm}

\begin{equation}
I(\phi)=\sum_{p=1}^{\infty} I_p \sin \left( 2\pi p \,\phi/\phi_0 \right)
\end{equation}

\vspace{1mm}

\noindent where the Fourier amplitudes $I_p$ are random variables drawn from probability distribution functions that are very nearly Gaussian \cite{smith92_gaussian1,eckern92_gaussian5,bussemaker97_gaussian2,houzet10_gaussian3,danon10_gaussian4}. (The randomness arises from the lack of control over the microscopic details of the ring). Each of these distributions is characterized by a mean value $\langle I_p \rangle$ and a width $\langle I_p^2 \rangle ^{1/2}$.

For non-interacting electrons, the mean values $\langle I_p \rangle$ are predicted to be very small \cite{altshuler91_prl66_1,schmid91_prl66_2,vonoppen91_prl66_3} (well below the sensitivity of existing instruments). As a result, the current in a particular ring will have Fourier amplitudes with random signs and magnitudes of the order of $ \langle I_p^2 \rangle ^{1/2}$, and the total signal from an array of $N \gg 1$ rings will be proportional to $N^{1/2}$. The  $\langle I_p^2 \rangle ^{1/2}$ decrease rapidly with increasing $p$, 
so $I(\phi)$ tends to be dominated by the $p = 1$ term \cite{riedel_vonoppen93,ginossar}. Thus, the signal from an array of $N \gg 1$ rings will have period $h/e$ and amplitude  $N^{1/2} \langle I_1^2 \rangle ^{1/2}$. The magnitude of the typical current $\langle I_p^2\rangle^{1/2}$ is set by the Thouless energy, $E_c=\hbar D/L^2$, where $D$ is the diffusion constant and $L$ the ring circumference. At low temperature ($T<T_c$, where $T_c=E_c/k_B$), $\langle I_1^2 \rangle ^{1/2} \sim eE_c/\hbar$ \cite{cheung_riedel_gefen_prl62_89,riedel_vonoppen93}.

For interacting electrons calculations predict that the average value of the second Fourier amplitude is greatly enhanced \cite{ambegaokar_eckern,altshuler_aronov,eckern}. Specifically, Ambegaokar and Eckern found that a Hartree-Fock treatment including a screened Coulomb interaction yields $\langle I_2 \rangle = \lambda e E_c/\hbar$ (for $T\ll T_c$) with the effective electron-electron coupling constant $|\lambda| \sim 0.1$ \cite{ambegaokar_eckern}. For repulsive electron-electron interactions $\lambda>0$, corresponding to paramagnetic susceptibility around $\phi = 0$. For attractive electron-electron interactions (such as in a superconductor above the critical temperature) $\lambda<0$, corresponding to diamagnetic susceptibility around $\phi = 0$ \cite{ae_euphyslett,von oppen_riedel,ambegaokar_eckern_d}. Thus for interacting electrons, the signal from an array of $N \gg 1$ rings is expected to be dominated by oscillations with period $h/2e$ and amplitude $ N \langle I_2 \rangle$. We emphasize that the interactions considered here are between the electrons in an individual ring. Interactions between rings in the array are assumed to be negligible.

This interaction-based enhancement of $\langle I_2 \rangle$ is predicted to be suppressed by modest magnetic fields \cite{ginossar}. Specifically, when the magnetic flux through the metal of the ring $\phi_M$ (distinct from the flux $\phi$ through the hole in the ring) satisfies $\phi_M  \gg \phi_0$, the enhancement vanishes and the persistent current is expected to be described by the non-interacting model.

To date, several measurements on arrays of rings \cite{levy1990,reulet1995,webb2001,deblock2002} have been interpreted as showing a diamagnetic persistent current with period $h/2e$ and an amplitude that is comparable to (but somewhat larger than) $N \langle I_2 \rangle$ as predicted by  Ambegaokar and Eckern \cite{ambegaokar_eckern}. These results have been obtained on arrays of rings made of Au \cite{webb2001}, GaAs \cite{reulet1995}, and Ag \cite{deblock2002}. Measurements of Cu rings found similar signals, but did not determine the sign \cite{levy1990,levy_physicab}. Interpreting these results within the Ambegaokar and Eckern theory would require the electron-electron interactions in these materials to be attractive. This is in apparent disagreement with the absence of a superconducting phase in these materials, as well as with other measurements (for a review see \cite{altshuler}).

Some possible explanations for this apparent discrepancy have been proposed. These  include a non-zero (but very low) superconducting transition temperature for some of these materials which is further suppressed by magnetic impurities within the ring \cite{bary-soroker_prl,bary-soroker_prb}, as well as  nonequilibrium effects due to high-frequency electromagnetic fields \cite{kravtsov_yudson,kravtsov_altshuler}. However to date there is not a complete understanding of what role these effects might play in the measurements of $\langle I_2\rangle$ reported in Refs. \cite{levy1990,webb2001,reulet1995,deblock2002}.

Here we describe the fabrication and characterization of a new type of device for measuring $\langle I_2 \rangle$. Each device consists of a large array of Au rings supported on a micromechanical cantilever. In the presence of an applied magnetic field persistent current in the rings will experience a torque, which in turn can be sensed as a static displacement of the cantilever or as a shift in the cantilever's resonant frequency. These devices will complement in a number of ways the SQUID-based and microwave-resonator-based detectors used in previous measurements of $\langle I_2 \rangle$ \cite{levy1990,webb2001,reulet1995,deblock2002}. First, they will measure persistent current (PC) in the presence of a somewhat larger magnetic field. The magnetic field must be large enough to generate a measurable torque, but small enough that $\phi_M$ is not much larger than $\phi_0$; as described below, this will correspond to 0.05 T $< B < $ 0.25 T. In contrast, previous measurements of $\langle I_2 \rangle$ were made with $B <$ 0.01 T. Second, the mechanical detectors described here will not produce high-frequency electromagnetic fields. This is in contrast to previous experiments, in which high-frequency fields were produced either as part of the measurement \cite{reulet1995,deblock2002} or as a consequence of Josephson oscillations in the SQUID junctions \cite{levy1990,webb2001}.

Mechanical detectors similar to the ones described here were used previously to measure the persistent current in Al rings in large $B$ \cite{ania_science,manuel13}. The large $B$ used in Refs. \cite{ania_science,manuel13} completely suppressed the superconductivity of the Al, and resulted in torsional signals with large signal-to-noise ratio (even for samples consisting of a single ring \cite{manuel13}). However the large $B$ ensured that $\phi_M \gg \phi_0$, and so precluded the observation of the interaction effects predicted in Refs. \cite{ambegaokar_eckern,ambegaokar_eckern_d,bary-soroker_prb,bary-soroker_prl}.
The devices described here incorporate a number of changes from those used in Refs. \cite{ania_vacsci,ania_science,manuel13}. First, the number of rings on a cantilever is increased by roughly three orders of magnitude to
 $\sim 10^5$ rings. Second, the rings' dimensions are chosen to minimize $\phi_M$ for a given $B$. Finally, the Al that was used in Refs. \cite{ania_vacsci,ania_science,manuel13} (and which is a superconductor at low $B$) is replaced by Au. To avoid proximity effects from commonly-used sticking layers (many of which are superconductors), the Au rings described here are deposited directly onto a silicon wafer.

\section{Sample fabrication}

\begin{figure}[h]
\vspace{2mm} \centerline{\hbox{
\epsfig{figure=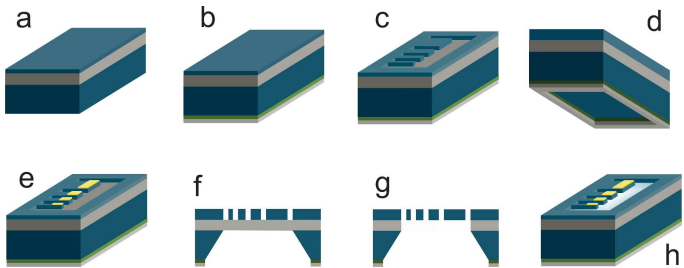,width=90mm}}}
\caption{A schematic illustration of the main fabrication steps. a) Starting SOI wafer. b) Silicon nitride and silicon oxide layers are deposited on the handle layer. c) Cantilevers are patterned on the front side. d) The etch mask out of silicon nitride and oxide is fabricated (view from below). e) Gold rings are evaporated on the cantilevers. f) KOH etch of the silicon handle opens the window to the cantilevers still standing on silicon oxide (cross-section). g) Silicon oxide is removed with HF (cross-section) leaving the cantilevers suspended. h) Fabricated sample with suspended cantilevers. Resist layers are not shown for simplicity.
}
\label{fabstages}
\end{figure}

The process starts from a $\langle 100\rangle$ silicon-on-insulator (SOI) 4-inch wafer (see Fig. \ref{fabstages}a). The 120 nm thick silicon device layer will eventually form the cantilevers after the underlying 400 nm thick BOx (buried silicon oxide) and 400 $\mu$m thick silicon handle layers are etched away. A layer of silicon nitride (200 nm) followed by a layer of silicon oxide (750 nm) is deposited by PECVD (plasma-enhanced chemical vapor deposition) on the handle layer (Fig. \ref{fabstages}b). These layers will be patterned into an etch mask. (The use of a bilayer provides stress compensation.)

To pattern the cantilevers, standard photolithograpy is used to define a photoresist mask on the device layer, followed by a CF$_4$ Reactive Ion Etch (RIE) (Fig. \ref{fabstages}c). Then a second photolithographic mask is defined on the back side of the wafer. This mask is aligned with the cantilever pattern on the front side of the wafer. The pattern is transferred to the oxide and nitride layers by a CHF$_3$/O$_2$ RIE (Fig. \ref{fabstages}d). In later steps these layers will serve as a hard mask for a wet etch.

To produce the rings, a MMA-MAA/PMMA bi-layer with a 270 nm/140 nm aspect ratio is spun on the device layer and exposed by an electron beam. The pattern is developed in MIBK/IPA at 25$^\circ$C for 50 s. 50 nm of 6N-purity gold is then deposited in an evaporator which has never evaporated magnetic materials. Lift-off is done in NMP at 80$^\circ$C (Fig. \ref{fabstages}e).

\begin{figure}[h]
\vspace{2mm} \centerline{\hbox{
\epsfig{figure=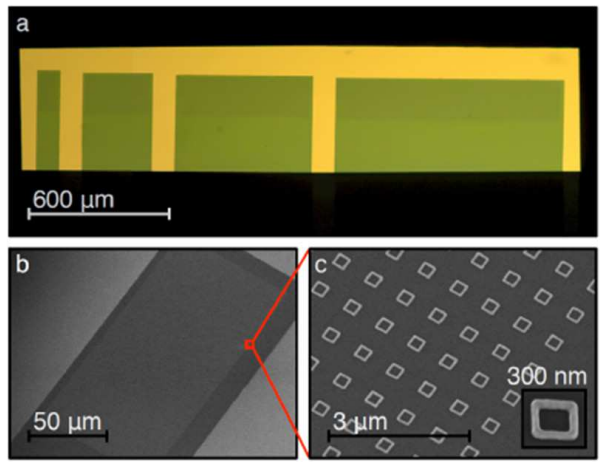,width=80mm}}}
\caption{a) Optical microscope image of the sample, where four cantilevers (green) of different width are fully suspended. The shaded area on the top of each cantilever is an array of closely packed gold rings. b) SEM photo of the top of a single cantilever resting on SiO$_2$. c) Zoom on the previous SEM photo showing the gold ring array. Inset is a close up of a single ring. Inset frame also serves as scale, showing 300 nm.
}
\label{samplephoto}
\end{figure}

To protect the gold rings during the wet etch of the handle layer, they are covered with KOH-resistant Pro-Tek B3 coating and the wafer is mounted on a specially made chuck which provides leak-tight coverage of the top side of the wafer. The silicon handle is then etched in a 30$\%$ water solution of KOH at 85$^\circ$C for approximately 4 hours at the rate of $\sim 100$ $\mu$m/h (Fig. \ref{fabstages}f). As the etch rate depends on the etchant concentration and temperature, and varies for different crystalographic axes of silicon, it is important to make sure that the solution is homogeneous (by constant stirring) and temperature stable (controlled by a feedback loop on the heater) throughout. The etch is quenched in water once the the BOx etch-stop layer is reached. The Pro-Tek layer covering the front side of the wafer was found to be important in preventing the KOH from passing through small cracks in the BOx layer.

The last step of the process is to release the cantilevers from the Pro-Tek film and BOx etch-stop layer. The SiO$_2$ layer is removed by immersing the chips in a 6:1 H$_2$O:HF solution with surfactant OHS (Fig. \ref{fabstages}g). The Pro-Tek film is then removed in NMP heated to 80$^\circ$C. The chips are transferred in fluid from NMP to acetone, methanol and finally isopropanol and critical point dried (Fig. \ref{fabstages}h). Occasionally the Pro-Tek layer can leave small amounts of residue on the cantilevers which are easily removed by a short CF$_4$/O$_2$ (5:45 sccm) RIE clean after the cantilevers are fully suspended.

Optical and SEM images of fabricated devices can be seen in Figure \ref{samplephoto}. The cantilevers (green) are all about 500 $\mu$m long and 100 $\mu$m to 1 mm wide. The rings are in fact squares as this shape is faster to pattern with the e-beam writer. The square sides are 250 nm long  and 50 nm wide. The spacing between adjacent squares' sides is 400 nm.

\section{Characterization of the cantilevers' mechanical properties}

The sample is placed in the vacuum space of a $^3$He fridge with 300 mK base temperature and an 8T superconducting magnet. The cantilever deflection is detected with a fiber-optic interferometer \cite{rugar,ania_science} using a 1550 nm laser. The interferometer signal is used as an input to a phase-locked loop, which in turn drives the cantilever at its resonant frequency (via a piezoelectric element). The frequency of this drive is monitored, and is used to infer the persistent current in the rings: as described in Ref. \cite{ania_science}, the change of the cantilever resonant frequency as function of field is proportional to $\partial I/\partial \phi$.

\begin{figure}[h]
\vspace{2mm} \centerline{\hbox{
\epsfig{figure=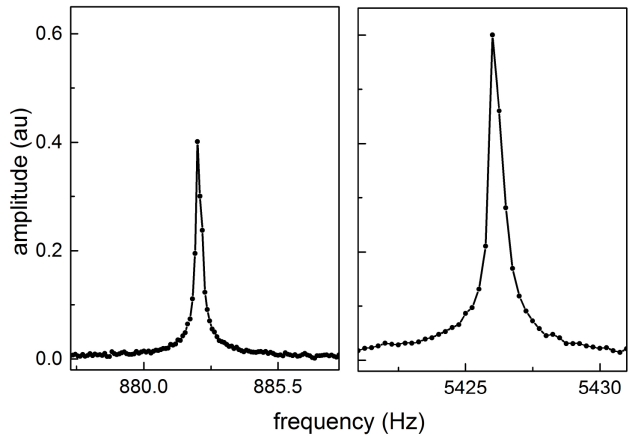,width=86mm}}}
\caption{Amplitude of the cantilever motion as function of drive frequency showing the first and second flexural mode, measured at $T=4$ K and $B=0$ T.}
\label{freqscan}
\end{figure}

Figure \ref{freqscan} shows a measurement of two resonances of a cantilever that is 500 $\mu$m long, 300 $\mu$m wide and 120 nm thick. The cantilever supports an array of $10^5$ rings. This measurement is performed by applying a sinusoidal drive to the piezoelectric element and then using the interferometer to record the amplitude of the cantilever's motion. The first flexural mode is at 882 Hz and the second at 5425 Hz. The first torsional mode (not shown) is at 4552 Hz.

\begin{figure}[h]
\vspace{2mm} \centerline{\hbox{
\epsfig{figure=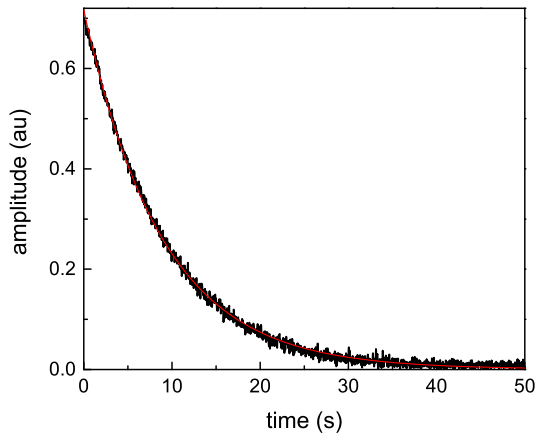,width=86mm}}}
\caption{Ringdown time of the resonance at the frequency $f_0=$ 882 Hz, measured at $T=370$ mK and $B=6$ T. Exponential fit is given in red, with the decay time $\tau=8.8$ s, yielding the quality factor $Q=\pi \tau f_0 \sim  2.5 \times 10^4$.}
\label{cantilever_ringdown}
\end{figure}

The quality factors $Q$ of each mode were determined by measuring the ringdown time. This measurement is performed by driving the cantilever on resonance, then turning off the drive and monitoring the amplitude of the cantilever's motion as function of time. A typical ringdown measurement at $T=370$ mK and $B=6$ T is shown in Figure \ref{cantilever_ringdown}. Fitting the data gives $Q \sim 2.5 \times 10^4$. Quality factors of other resonances were similar and they were found not to be strongly affected by the magnetic field.

\section{Expected signal}

To estimate the likely signal from $\langle I_2 \rangle$, we note that previous experiments on ensembles of rings have found amplitudes $0.3-0.77 eE_c/\hbar$ \cite{levy1990,webb2001,reulet1995,deblock2002}. Assuming that $\langle I_2 \rangle=0.5e E_c/\hbar$ in the Au rings described above, we expect this to correspond to 1 nA in each ring. This estimate is based on the transport measurements (not shown) of wires deposited from the same source material, which give $D=140$ cm$^2$/s. The estimate $\langle I_2 \rangle=1$ nA includes the effect of the expected measurement temperature ($T=300$ mK), which results in a $50\%$ reduction of $\langle I_2 \rangle$ from its $T=0$ value.

The magnetic moment associated with the persistent current $I(\phi)$ shifts the cantilever resonant frequency by an amount

\begin{equation}
\Delta f =N \frac{f_0}{2k}\left(\frac{\alpha}{l}\right)^2\left(A B \sin \theta_0 \right)^2\frac{\partial I}{\partial \phi}
\label{eq1}
\end{equation}

\noindent where $f_0$ is the cantilever resonant frequency, $k$ the spring constant, $l$ the cantilever length, $\alpha$ the geometrical factor of the order of 1 related to the mode-shape, $A$ the surface of the ring, $B$ the external magnetic field and $\theta_0=\pi/4$ the angle between the field and the perpendicular to the ring plane \cite{ania_science}. Taking $\partial I/\partial \phi \sim I/\phi_0$, the current of 1 nA in $N=10^5$ rings  should produce a  frequency shift of $\Delta f=15$ $\mu$Hz.

\begin{figure}[h]
\vspace{2mm} \centerline{\hbox{
\epsfig{figure=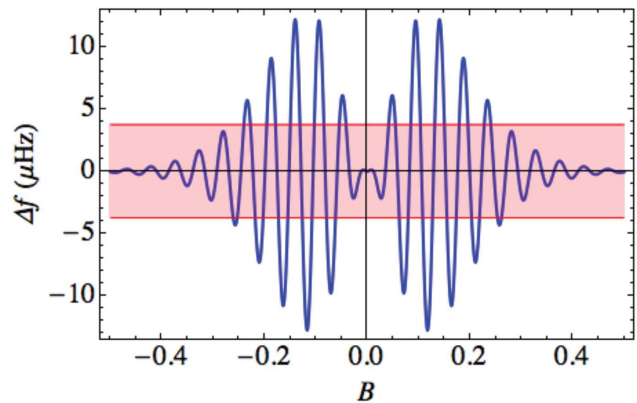,width=86mm}}}
\caption{Expected shift of the cantilever resonant frequency as function of field, calculated from \cite{bary-soroker_prb,bary-soroker_prl}.}
\label{expectedsignal}
\end{figure}

Figure \ref{expectedsignal} shows the expected frequency shift in the field range of interest. The signal scales as $B^2$ for small fields (Eq. \ref{eq1}) and then decays at higher fields for which $\phi>\phi_M$ (here $\sim$ 0.2 T) due to the suppression of the interaction current by flux penetrating the metal of the ring. This suppression is calculated following Ref. \cite{ginossar}. The sign of the current corresponds to the sign of the frequency shift such that $I \rightarrow -I$ would lead to $\Delta f \rightarrow -\Delta f$.

The sensitivity of similar measurements is typically limited by the thermal motion of the cantilever. This motion sets a minimum detectable frequency shift

\begin{equation}
(\delta f)^2 = \frac{1}{2 \tau_M} \frac{ f_0 k_B T}{\pi Q k x_0^2}
\end{equation}

\noindent where $\tau_M$ is the measurement time and $x_0$ the maximum displacement of the cantilever \cite{ania_vacsci}. If we take $\tau_M=100$ s and $x_0=200$ nm (typical values for similar measurements \cite{ania_science,manuel13}) we get $\delta f \sim $ 4 $\mu$Hz at 300 mK, shown as the noise floor on Fig. \ref{expectedsignal}. As a result, we expect that these devices will allow for measurement of $\langle I_2 \rangle$, and may provide insight into the role of electron-electron interactions in persistent current and weak localization in thermal equilibrium.

We are grateful to Hugues Pothier for assistance in fabricating the Au
samples. We would like to acknowledge support from the National Science Foundation (NSF) (Grant No. 1106110) and from the US-Israel Binational Science Foundation (BSF). Facilities use was supported by YINQE.

\end{document}